\documentclass[a4paper,11pt]{article}
\usepackage{pos}
\usepackage{graphicx} 
\usepackage{url}

\begin{document}

\title{The Schwarzschild-Couder Telescope for the Cherenkov Telescope Array Observatory}

\author*{Luca Riitano}
\author{Justin Vandenbroucke}
\author{Zach Curtis-Ginsberg}
\affiliation{Department of Physics, University of Wisconsin-Madison,\\1150 University Avenue,Madison,USA}
\affiliation{For the SCT Collaboration and \href{https://www.ctao.org}{CTAO Consortium}}
\emailAdd{riitano@wisc.edu}

\abstract{The Cherenkov Telescope Array Observatory (CTAO) will greatly improve upon sensitivities in the field of very-high-energy gamma-ray astrophysics. The CTAO northern site (CTAO-North, La Palma, Spain) currently hosts LST-1 with the remaining three large-sized telescopes (LSTs) expected in mid-2026 and one medium-sized telescope (MST) expected in mid-2027. The CTAO southern site (CTAO-South, Paranal, Chile) expects the delivery of five small-sized telescopes (SSTs) and two MSTs in early 2026 with on-site construction beginning in mid-2026. The dual-mirrored Schwarzschild-Couder Telescope (SCT) is a candidate MST for CTAO-South and is capable of observations in the energy range of 100 GeV to 10 TeV, the core of CTAO’s 20 GeV to 300 TeV energy range. Inaugurated in January 2019, the prototype SCT (pSCT) located at the Fred Lawrence Whipple Observatory in southern Arizona observed gamma-ray emission from the Crab Nebula at a significance of 8.6 sigma in 2020. The pSCT utilizes a novel dual-mirror optics design and a densely packed focal plane of silicon photomultipliers (SiPMs). An upgrade of the pSCT camera is underway to fully instrument the camera with 11,328 pixels and an 8-degree diameter FoV. In addition, upgraded electronics will lower the front-end electronics noise, allowing for a lower trigger threshold and improved event reconstruction and background rejection. This work will present the status of the upgrade of the pSCT and discuss the future of the SCT.}

\FullConference{39th International Cosmic Ray Conference (ICRC 2025)\\
CICG - International Conference Centre - Geneva, Switzerland\\
14 - 24 July, 2025
}
\ConferenceLogo{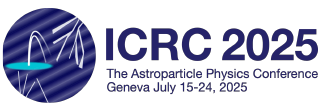}

\maketitle

\section{Introduction}
When very-high-energy (VHE) gamma rays strike the Earth's atmosphere, they initiate extensive air showers (EAS) of neutral and charged particles. The charged particles, moving faster than the speed of light in the medium, produce Cherenkov radiation that is detectable by Imaging Atmospheric Cherenkov Telescopes (IACTs). CTAO will improve upon the sensitivity of the current generation of IACTs with two sites, offering full sky coverage at energies of 20 GeV to 300 TeV.

\begin{figure}[htp]
    \centering
    \includegraphics[width=0.9\textwidth]{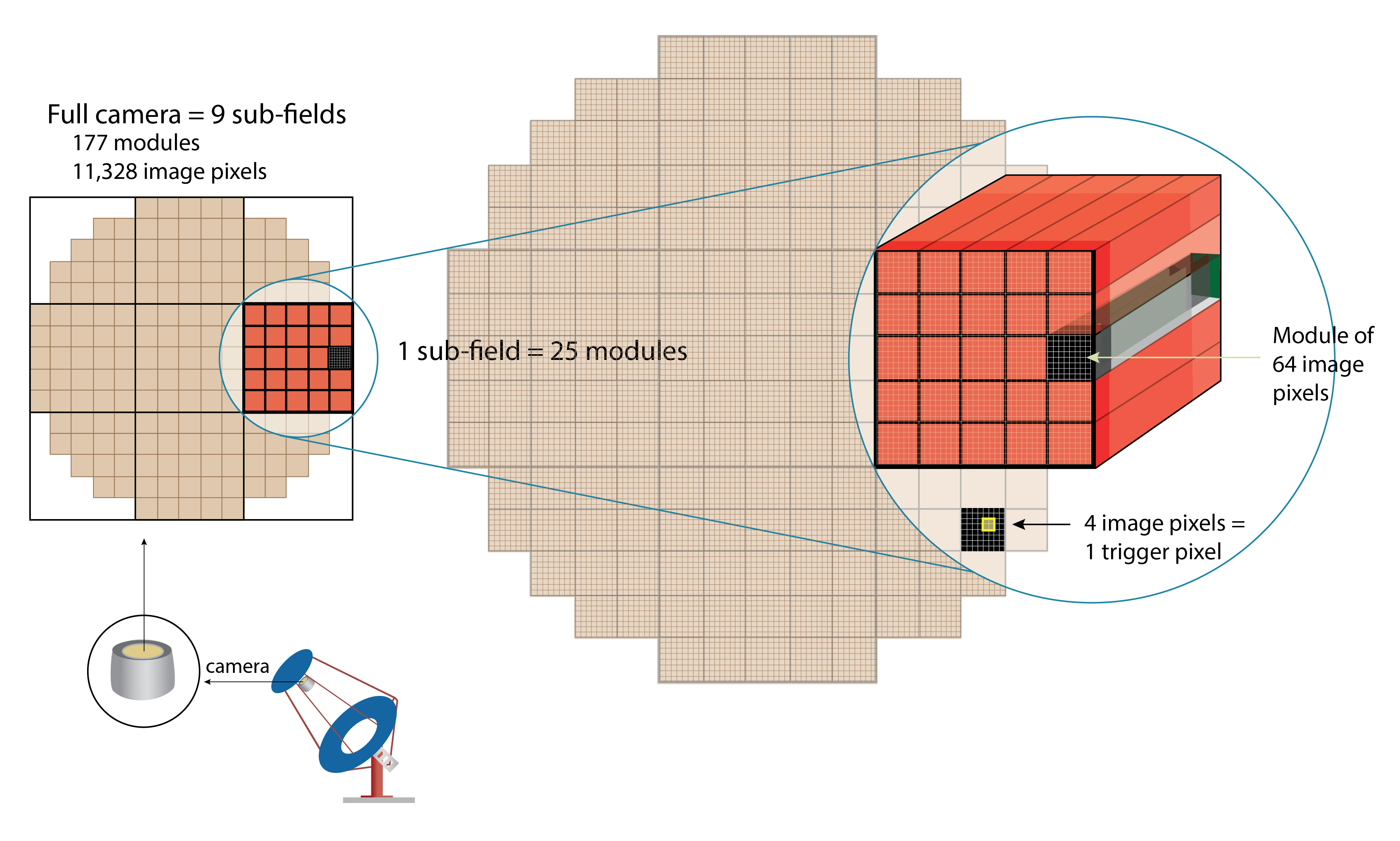}
    \caption{Camera mapping of the SCT design. The SCT has nine sectors, with the four corner sectors containing 13 modules each, and the other 5 sectors containing 25 modules each. Every module houses 16 trigger pixels - formed by four image pixels - and 64 image pixels.}
    \label{fig:SCT_cam_design}
\end{figure}

The SCT is a candidate MST design for CTAO-South with sensitivity in CTAO's core energy range of 100 GeV to 10 TeV and a large 8 degree wide field of view (FoV). The design implements Schwarzschild-Couder (SC) dual-mirror optics as opposed to the more common Davies-Cotton (DC) single-mirror optics. The SC optical design reduces optical aberrations and achieves near isochronism, leading to an improved point spread function (PSF) and reduced trigger coincidence window. The 9.66 m primary and 5.4 m secondary mirror focus images to a reduced plate scale of 1.625 mm per arcmin. The small plate scale is exploited more effectively by implementing compact 6 mm square SiPM pixels as opposed to the more commonly employed photomultiplier tubes (PMTs); the SCT design uses many times more pixels than any other CTAO design: 11,328 pixels grouped in 177 modules and 9 sectors (figure \ref{fig:SCT_cam_design}) \cite{Vassiliev_2007}.

\begin{figure}[htp]
    \centering
    \includegraphics[width=0.45\textwidth]{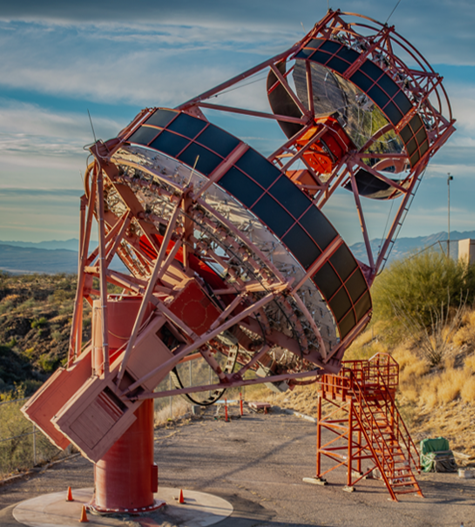}
    \caption{Photograph of the pSCT at Fred Lawrence Whipple Observatory in Arizona with the old camera installed.}
    \label{fig:pSCT}
\end{figure}

\begin{figure}[htp]
    \centering
    \includegraphics[width=0.55\textwidth]{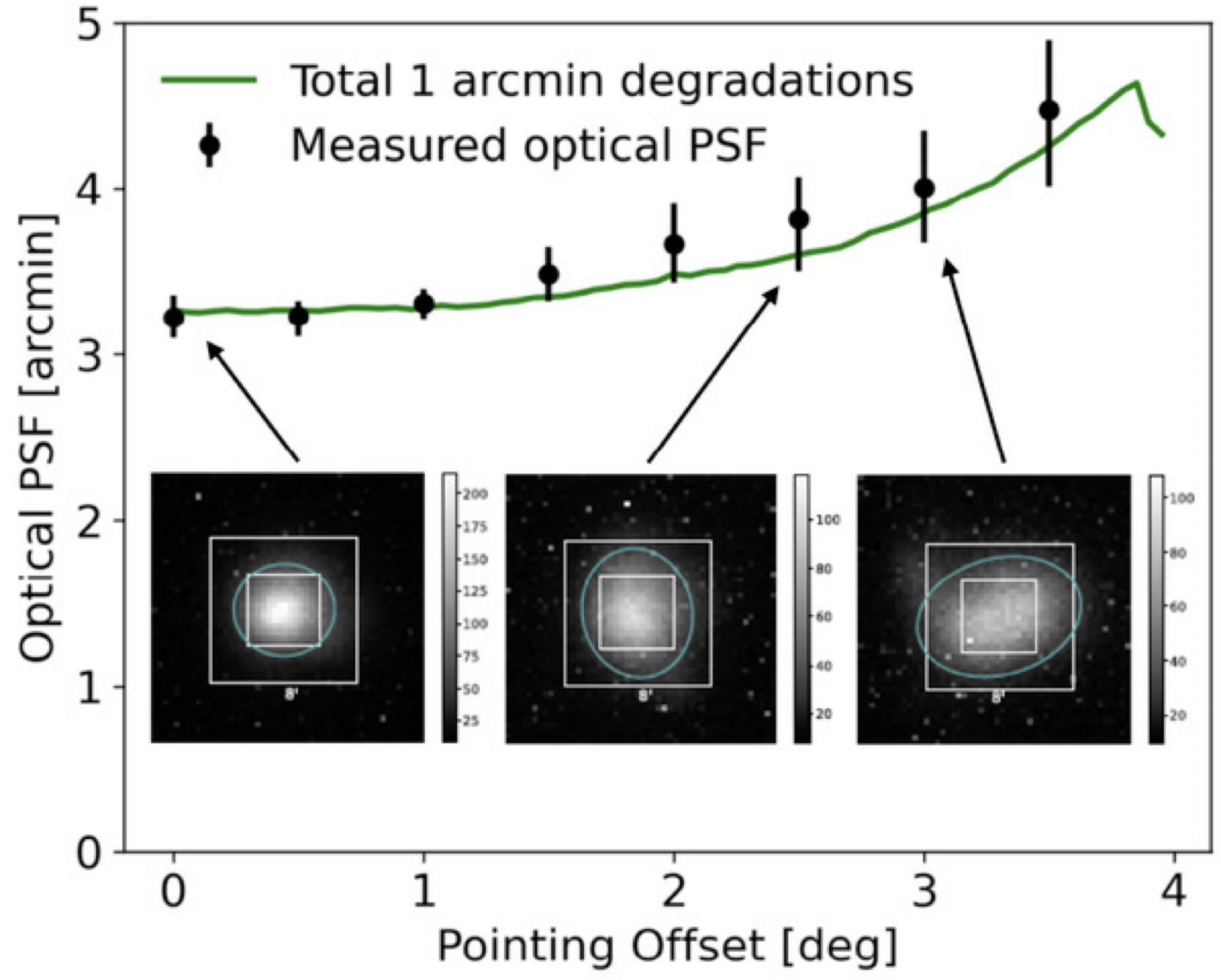}
    \caption{Optical PSF of the pSCT as a function of pointing offset, as measured using a bright star. The PSF is calculated by fitting a 2-dimensional Gaussian to the image and taking twice the value of the larger fit sigma.}
    \label{fig:PSF}
\end{figure}

The pSCT (pictured in figure \ref{fig:pSCT}) was inaugurated in January 2019 at Fred Lawrence Whipple Observatory with a partially populated focal plane. Of the 177 total modules, only the 25 modules (1600 pixels) of the central sector were produced, resulting in a 2.7 degree wide FoV \cite{camera_paper_jatis}. The mirrors were aligned to achieve an optical PSF of approximately 3 arcminutes on-axis, degrading only to 4 or 5 arcminutes for off-axis measurements (figure \ref{fig:PSF}). Observations of the Crab Nebula in 2020 culminated in a detection of gamma-ray emission at a significance of 8.6 sigma \cite{ADAMS2021}.

\section{Camera Upgrade}

The pSCT camera upgrade includes updates to both the camera mechanical design and the module electronics \cite{taylor2022}. The upgraded camera structure was installed in March of 2024 with upgrades to the cooling system, mechanical motion system, and power dissipation (see \cite{Kieda_2025}). A few early upgrade camera modules have been produced and tested in preparation for mass production and calibration.

\begin{figure}[htp]
    \centering
    \includegraphics[width=0.8\textwidth]{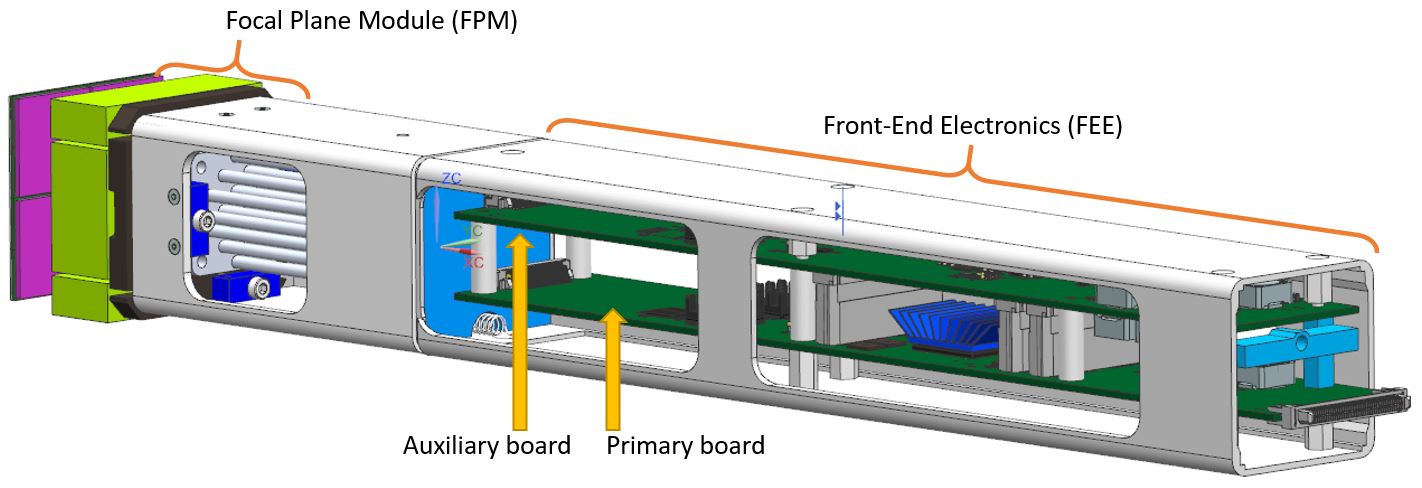}
    \caption{Rendering of the upgrade CTC module. At the far left, the SiPMs are visible without the SMARTs attached to the back. The two boards of the FEE inside their cage are shown on the right.}
    \label{fig:module_drawing}
\end{figure}

The upgraded pSCT module (figure \ref{fig:module_drawing}) is divided into a focal-plane module (FPM) and front-end electronics (FEE). The FPM consists of several components including SiPM pixels and SiPM Multichannel ASICs for high-Resolution cherenkov Telescopes (SMARTs). The SiPMs are mounted on a copper post that is in thermal contact with a Peltier element maintaining a user-input temperature. The main components of the FEE are the TeV Array Readout GSa/s sampling and Event Trigger (TARGET) application specific integrated circuits (ASICs). The two ASIC designs used in the FEE are Cherenkov TARGET-C (CTC) for sampling and digitization and the Cherenkov TARGET-5 Trigger Extension ASIC (CT5TEA) for triggering. The combination of ASIC designs replaces the single TARGET-7 (T7) ASIC design used in the camera during initial operations. The division of triggering and digitization electronics into multiple ASICs will reduce the data path noise. More details of the TARGET ASICs are reported in \cite{Funk_2017, Schwab_2024}. The module is divided into four 16 channel quadrants that each make use of their own set of SMART, CT5TEA, and CTC ASIC. The FEE is comprised of two boards that are responsible for two quadrants each - a primary board with a Xilinx Artix-7 XC7A100T-FGG484 field-programmable gate array (FPGA) and an auxiliary board.

The upgrade modules use Fondazione Bruno Kessler (FBK) Near Ultra-Violet High-Definition (NUV-HD) SiPMs with high photon detection efficiencies of up to approximately 50\% (see \cite{AMBROSI2023, Merzi_2023, LOIZZO2024}). Directly connected to the SiPMs are SMART v2 ASICs which amplify and shape the SiPM signals before transmission to the FEE. Additionally, SMART ASICs adjust the fine bias of the SiPMs via a digital-to-analog converter (DAC) and monitor the SiPM currents (see \cite{ARAMO2023_1045, ARAMO2023_1047, ARAMO2023_IWASI}).

The CT5TEA ASIC is responsible for the low-level trigger and controls the readout voltage baseline via a 12-bit DAC. Groups of four adjacent data channels are combined to form each trigger pixel. The sum of the data signals is compared to an adjustable threshold to determine if a trigger signal should be generated. The CTC ASIC samples the signal transmitted by the FPM at 1 GSa/s and stores a charge proportional to the measured signal in a 16,384 sample ring buffer switched-capacitor array. When the CTC receives a trigger signal, the charge within a user-defined readout window is readout out using Wilkinson ramp analog-to-digital converters (ADCs).

\section{Performance}

\begin{figure}[htp]
    \centering
    \includegraphics[width=0.8\textwidth]{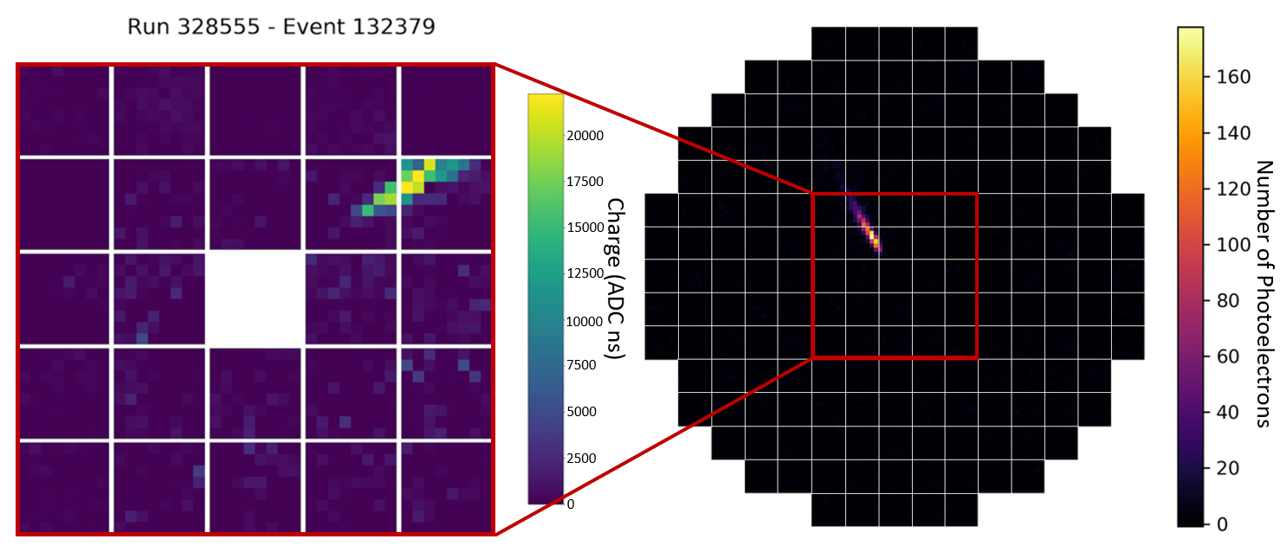}
    \caption{The first gamma ray detected by the pSCT on January 17, 2020 during observations of the Crab Nebula is shown on the left. The center module had been replaced with an optical alignment module, so the camera was operating with 24 camera modules. The event was confirmed as a gamma ray by timing coincidence with the VERITAS IACT array (colocated at FLWO). On the right is a simulated 2.72 TeV gamma ray in the upgraded pSCT camera.}
    \label{fig:gamma_ray_comparison}
\end{figure}

The lower electronics noise and increased FoV of the upgraded camera is expected to significantly improve upon the performance of the pSCT. A real gamma-ray image from observations of the Crab Nebula in 2020 is compared to a simulated gamma ray in the upgraded pSCT camera (figure \ref{fig:gamma_ray_comparison}).

The performance of a test camera module was evaluated in the lab using a custom LED flasher. For each data run, the flasher and module were triggered synchronously via a function generator. The data was calibrated and converted from ADC counts to millivolts before the charge was calculated as a sum of samples over an idealized 9 ns window. The resulting charge was then binned into a histogram, sometimes called a "finger plot", so that individual photoelectron peaks could be fit to extract parameters that define the camera module's performance.

\begin{figure}[htp]
    \centering
    \includegraphics[width=0.7\textwidth]{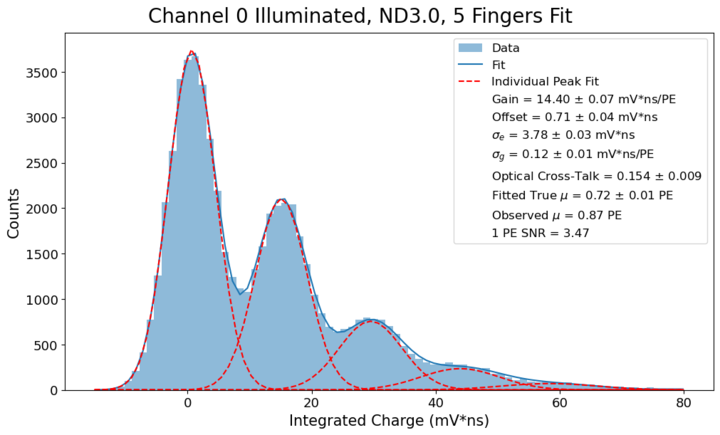}
    \caption{Histogram of calculated charges in channel 0 for a run of approximately 60,000 low intensity LED flasher events. A fully constrained fit was performed as a sum of Gaussians. The red lines indicate the individual Gaussians while the light blue line indicates the sum of the Gaussians. The final fit parameters are included in the legend.}
    \label{fig:finger_plot_fully_constrained}
\end{figure}

The individual photoelectron peaks in the histogram can be modeled by Gaussian fits. Each Gaussian has a mean equal to the channel gain multiplied by the number of photoelectrons. The variance of each Gaussian is the sum in quadrature of the standard deviations of two contributing sources of noise: the electronics noise and the gain noise. The electronics noise is constant for each Gaussian but the gain noise scales with the mean of the Gaussian. To account for the waveform baseline shifts that can occur due to temperature fluctuations between data and calibration runs, an offset parameter is also included that shifts the Gaussian means by a constant value. The relative amplitude of each Gaussian can also be constrained using a generalized Poisson distribution described in \cite{VINOGRADOV2012}. The generalized Poisson distribution determines the relative number of photoelectrons in each Gaussian through two parameters: the optical crosstalk fraction and the mean number of photoelectrons detected by the SiPM pixel per event (before optical crosstalk). By fitting the sum of these Gaussians, the six aforementioned parameters can be extracted. An example of this fit was performed in figure \ref{fig:finger_plot_fully_constrained} and the signal-to-noise ratio (Gaussian mean divided by standard deviation) calculated for the single photoelectron peak. Medium charge flasher runs which cannot be easily fit with Gaussians can still be fit with the generalized Poisson distribution to extract a gain, true mean charge, and optical crosstalk value.

\begin{figure}[htp]
    \centering
    \includegraphics[width=0.7\textwidth]{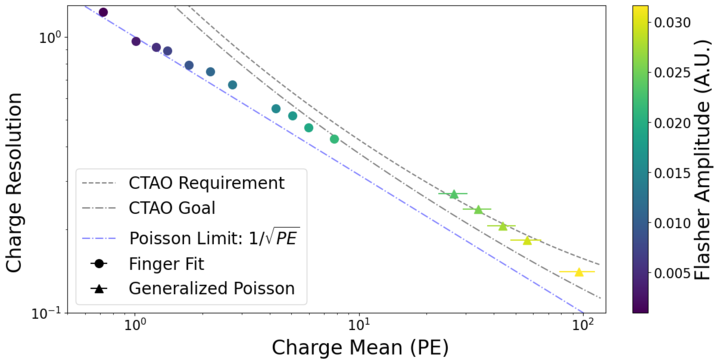}
    \caption{Charge resolution of 16 low to medium intensity LED flasher runs as a function of mean charge. The circles indicate runs with fully constrained finger plot fits while the triangles indicate runs with simple generalized Poisson fits. The color of the marker denotes the relative flasher intensity. The light blue dash-dot line indicates the physical limit to charge resolution set by the Poisson limit, while the other two gray lines indicate the former CTAO charge resolution requirement and goal.}
    \label{fig:charge_resolution}
\end{figure}

The charge resolution - defined as the standard deviation of the charge divided by the mean of the charge in a single run - can be calculated at different intensities for each channel. The charge resolution is theoretically limited by the Poisson limit to be \begin{math}1/\sqrt{NPE}\end{math} where NPE is the mean number of photoelectrons per event. The other sources of uncertainty that worsen the charge resolution are the electronics noise, a constant, and the gain uncertainty, which is proportional to \begin{math}1/NPE\end{math}. CTAO previously released a charge resolution requirement and goal for CTAO telescopes, \begin{math}\sqrt{0.1^2 + 1.2NPE^{-\frac{1}{2}} + 1.6NPE^{-1}}\end{math} and \begin{math}\sqrt{0.05^2 + 1.1NPE^{-\frac{1}{2}} + 1.4NPE^{-1}}\end{math}, respectively, that we can use to evaluate our camera module performance. The charge resolution for low and medium charge runs is plotted against the Poisson limit and CTAO benchmarks in figure \ref{fig:charge_resolution}. The low charge finger plot points, with intensities of around 1-10 photoelectrons, all fall below the CTAO goal, with one point falling below the Poisson limit (likely due to unaccounted for systematic uncertainties). The medium charge, generalized Poisson runs all fall below or within one standard deviation of the CTAO goal. The differences between the two sets of runs are likely due to the differences in systematic uncertainties of the fit methods.

\section{Conclusion and Outlook}

\begin{figure}[htp]
    \centering
    \includegraphics[width=0.8\textwidth]{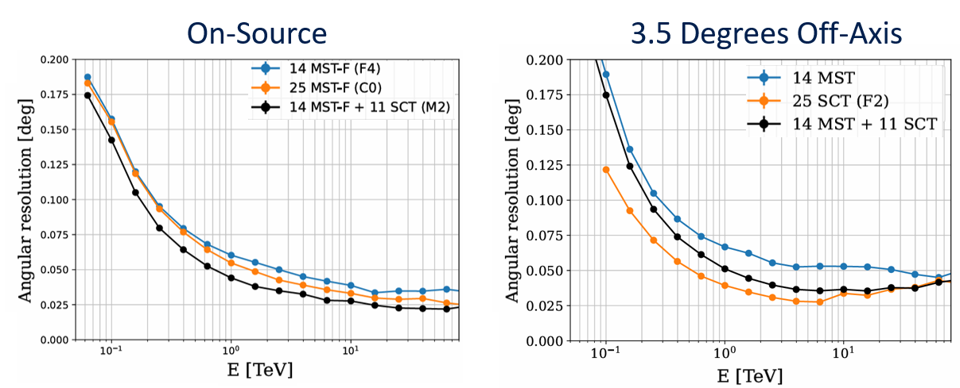}
    \caption{Angular resolution as a function of energy for different MST and SCT configurations. On the left is the angular resolution for on-source observations while the right plot displays the angular resolution for observations 3.5 degrees off-axis. The values were calculated using the CTAO prod3b IRFs.}
    \label{fig:SCT_angular_resolution}
\end{figure}

\begin{figure}[htp]
    \centering
    \includegraphics[width=0.8\textwidth]{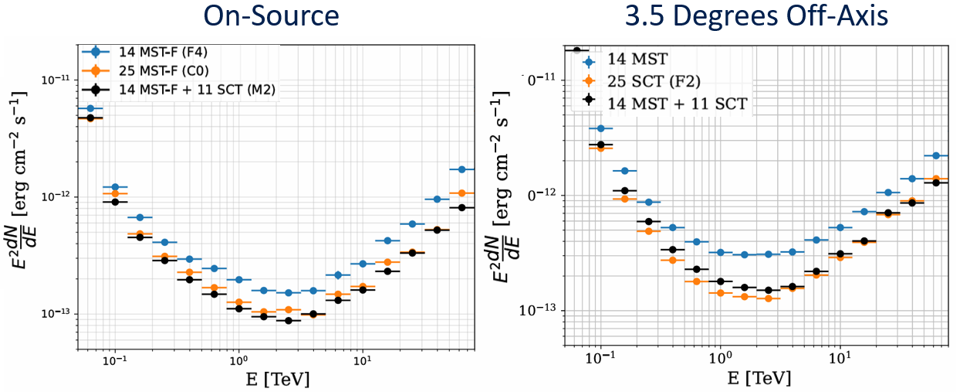}
    \caption{Flux sensitivity as a function of energy for different MST and SCT configurations. On the left is the sensitivity for on-source observations while the right plot displays the s for observations 3.5 degrees off-axis. The values were calculated using the CTAO prod3b IRFs.}
    \label{fig:SCT_sensitivity}
\end{figure}

The initial CTAO configuration will include 14 MSTs of the DC design and no SCTs. Simulations of the initial configuration of MSTs and the three potential upgrade configurations were performed using CTAO prod3b instrument response functions. The angular resolution was compared across configurations and found to improve at all energies for on-source observations as MSTs were exchanged for SCTs. The improvement was even stronger at 3.5 degrees off-axis (fgure \ref{fig:SCT_angular_resolution}). Similarly, the sensitivity of the configurations improved as MSTs were replaced with SCTs (figure \ref{fig:SCT_sensitivity}).

Through its detection of the Crab Nebula in 2020, the pSCT has already proven its scientific capabilities. The upgraded camera has been installed in the telescope and mass production of the upgrade modules is imminent. The upgrade camera module will lower the electronics noise, improving trigger performance and event reconstruction. The full population of the focal plane will expand the pSCT FoV to its full 8 degree diameter. With an upgraded camera, the pSCT will demonstrate the capabilities of a single SCT.

\bibliography{references}
\bibliographystyle{JHEP}

\end{document}